\documentclass{aa}
\usepackage{graphicx}

\begin{document}

\thesaurus{06    
             (08.16.4; 
              08.03.2; 
              08.03.4; 
              08.09.2: IRAS 03313$+$6058)}

\title{IRAS 03313$+$6058: an AGB star with 30 micron emission
\thanks{Based on observations with ISO, an ESA project with 
instruments funded by ESA Member States (especially the PI countries: 
France, Germany, the Netherlands and the United Kingdom) and with 
participation of ISAS and NASA}
}

\author{B. W. Jiang\inst{1,2}, R. Szczerba\inst{2,3}, \and S. Deguchi\inst{4}}

\offprints{Biwei Jiang (jiang@class1.bao.ac.cn)}

\institute{ 
          Beijing Astronomical Observatory, Chinese Academy of 
          Sciences, Datun Rd.  No.20(A), Beijing 100012, China 
\and 
          N. Copernicus Astronomical Center, ul.  Rabia\'{n}ska 8, PL-87--100 
          Toru\'{n}, Poland 
\and 
          Department of Physics and Astronomy, University of Calgary,
          Calgary, Alberta T2N 1N4, Canada
\and
          Nobeyama Radio Observatory, Minamimaki, 
          Minamisaku, Nagano 384--1305, Japan}

\date{Received date: 25 September 1998; accepted date: 16 December 1998 }
 
\authorrunning{B.W. Jiang et al.}
\maketitle

\begin{abstract} This paper reports a detection of the 30 micron 
emission feature from the C--rich Asymptotic Giant Branch (AGB) star IRAS 
03313$+$6058 based on the ISO--SWS observation.  Modeling of the 
spectral energy distribution shows that this emission starts at about 20 
micron and possibly extends to the limit (45 micron) of the observation.  
By assuming MgS as the carrier, the number ratio of Sulfur atom in MgS to 
Hydrogen atom in total, n(S)/n(H), is 3.0$\times$10$^{-6}$ from 
model fitting.  A comparison of this emission feature is made with 
other AGB and post--AGB objects.

\keywords{Stars: AGB and post--AGB  -- circumstellar matter -- stars: 
chemically peculiar -- Stars: individual: IRAS\,03313$+$6058}

\end{abstract}

\section{Introduction}
\indent

The 30 micron emission was first observed in the bright AGB star, 
IRC$+$10216 (Low et al 1973).  Observations of several 
carbon stars by Goebel \& Moseley (1985) found that the emission also
occurs in AFGL 3068. 
Recent ISO observations of some carbon stars 
turned up a few more candidates: AFGL\,2256, AFGL\,2155 and 
IRC$+$40540 (Yamamura et al. 1998). 
This feature is seen not only in extreme 
C--rich AGB stars, but also in C--rich proto--planetary nebulae (PPNe; 
called also post--AGB objects) and planetary nebulae (PNe).  Omont et 
al.  (1995) observed five C--rich PPN objects with an unidentified 
emission feature at 21 micron in their IRAS LRS spectra (Kwok et al.  
1989) and detected the 30 micron feature in all of them.  In the 
case of PNe a similar feature is observed for IC\,418 and NGC\,6752 
(Forrest et al.  1981).  A few suggestions for the carrier of this 
emission have been proposed.  Because the feature has never been detected 
in O--rich objects, and is the broadest known feature in the mid--infrared, 
solid species that form in the absence of oxides are taken into 
consideration.  Omont et al.  (1995) suggested iron atoms bound to PAH 
molecules as a possible emitter for the 30 micron feature.  But Chan 
et al.  (1997) raise doubt about this suggestion after the detection 
of this feature in direction of the Galactic center, 
where the existence of PAH 
molecules is questionable.  A more acceptable candidate is solid 
magnesium sulfide (MgS), first suggested by Goebel \& Moseley (1985).  
The laboratory spectra of MgS samples showed very good agreement of 
band turn--on and cut--off wavelengths, as well as the overall band shape, 
with 
the observed feature seen in AFGL\,3068 and IRC$+$10216.  A reasonable fit 
was achieved to the 30 micron feature of the PPN object IRAS\,22272$+$5435 
(Szczerba et al.  1997) by means of MgS grains with a distribution of 
shapes.  In 
addition, MgS is one of the molecules which condensate at low 
temperature when no oxides are present.  This is consistent with 
the detection of the 30 micron feature only in objects with cold 
dust shells.

IRAS\,03313$+$6058 was classified as a candidate extreme carbon star 
based on the similarity of its IRAS LRS spectrum to those of AFGL\,3068 and 
IRC$+$10216 (Volk et al.  1992).  It has no optical counterpart in the POSS 
plates (Jiang \& Hu 1992).  In the near infrared, the source was detected 
at K$=$15.6 mag and was not detected in J and H bands with upper limit of  
magnitude 17 and 16, respectively (Jiang et al. 1997).  
The color index between 12 and 25 micron, based on the IRAS PSC 
catalogue, indicates a color 
temperature of about 250\,K. Therefore, this is an 
object with a cold and optically 
thick circumstellar envelope.  Though the IRAS 
LRS type of this object is 22, no clear silicate feature is seen (note that 
Kwok et al.  1997 classified the LRS spectrum of this object as 
unusual).  The detection of HCN\,(1--0) line (Omont et al.  1993) 
indicated the possibility of a C--rich nature.   
O--rich maser lines such as OH (Le Squeren et al.  1992, Galt et al.  
1989), H$_{2}$O (Wouterloot et al.  1993) or SiO (Jiang et al.  1996) 
have not been detected at all and this 
is an indirect indication of a C--rich circumstellar envelope. The CO (2--1) 
line profile suggests that this object is 
a late--type star rather than a young stellar 
object, and gives an expansion velocity of the shell of 13.9\,km/s (Volk et 
al.  1993).

\section{Observation and data reduction}
\indent

The spectroscopic observation was carried out by using the SWS 
spectrometer (de Graauw et al.  1996) of the Infrared Space 
Observatory (ISO) satellite on 31 July 1997 with the fastest scan speed 
covering full wavelength range from 2.3 $\mu$m to 45 $\mu$m (AOT\,01, speed 
1).  The achieved resolution is about 200 $\sim$ 300 with S/N higher 
than 100 at wavelength longer than 5\,$\mu$m.

The original pipeline data were corrected for dark current, up--down 
scan difference, flat--field and flux calibration by using the SWS 
Interactive Analysis (IA) at MPE Garching \footnote{We acknowledge 
support from the ISO Spectrometer Data Centre at MPE Garching, funded 
by DARA under grant 50 QI 9402 3}.  The deglitching and averaging to 
equidistant spectral point in wavelength across the scans and detectors was 
done using the ISAP package.

The ISOCAM imaging was performed on 31 July 1997 with the CAM camera 
in the mode AOT\,01 with the filter LW3, centered at 15 micron.  The 
data were reduced by using CIA  (a joint development by the ESA 
Astrophysics Division and the ISOCAM Consortium led by the ISOCAM PI, 
C. Cesarsky, Direction des Sciences de la Matiere, C.E.A., France; 
Cesarsky et al 1996).  The object is still point-like
at this angular resolution of 1.5 arcsec/pixel.  The 
flux through the filter LW3 is 59.03 Jy as measured by aperture 
photometry method, about twice of the 12 $\mu$m flux (30.87\,Jy) given by the
IRAS PSC catalogue. This difference could be caused by the strong emission in 
the mid--infrared range of the spectrum or by the object being variable. 
However, the 
flux ($\sim$ 39 Jy) at 15 micron of the SWS spectrum, which was taken 
at the same day as the ISOCAM image, does not match the photometric result 
from the ISOCAM image while it is in rough agreement with the 12 $\mu$m IRAS
flux. By calculating the IRAS fluxes at 12 and 25 
micron from SWS, correction factors of 1.11 and 1.06 should be applied to 
the corresponding SWS bands, respectively, to agree with the photometric 
data from IRAS PSC. After 
correction, the spectrum is smooth and the shape is similar to its 
IRAS LRS spectrum.  
In the same time, the IRAS LRS spectrum should be 
multiplied by a factor of 1.27 to agree with the IRAS PSC data.
These factors are within the calibration 
uncertainties of the ISO--SWS and the IRAS LRS. Such agreement may show that
the flux calibration of ISO--SWS is reliable, and that the object is
non--variable (the variability index from IRAS observation is 50, at 
the border between variable and nonvariable indices).  
Therefore, we suspect that the discrepancy  between ISO--SWS and ISO--CAM 
results is related to the uncertainties in the flux calibration
from the image at LW3 band, but the reason for this is not clear to
us. Anyway this ISO-CAM result is shown in Fig. 1 as an open circle.

\section{Modeling}
\indent

\subsection{Spectral Energy Distribution}

The overall spectral energy distribution of this object is 
observationally determined from the near infrared to the far infrared.  The K 
band magnitude at 2.2 micron is 15.6 (Jiang et al.  1997), i.e., the flux is  
3.7$\times$10$^{-4}$\,Jy.  The fluxes at 12, 25, 60 and 100\,$\mu$m 
are 30.87, 43.44, 15.08 and 6.40\,Jy, respectively, from the IRAS 
photometric observations ( quality index = 3 in all four IRAS bands).  
Combination of these photometric results with the ISO SWS\,01 spectrum 
defines the observed spectral energy distribution.  In Fig.  1, the 
observational data are shown where the dots represent the photometric 
results and the thin solid line shows the ISO SWS\,01 spectrum.  The 
spectrum of ISO--SWS from 2.3 micron to 3.52 micron (band 1A, 1B and 
1D of ISO--SWS) is not shown because it is too noisy.

\begin{figure} 
\resizebox{\hsize}{!}{\includegraphics[angle=-90]{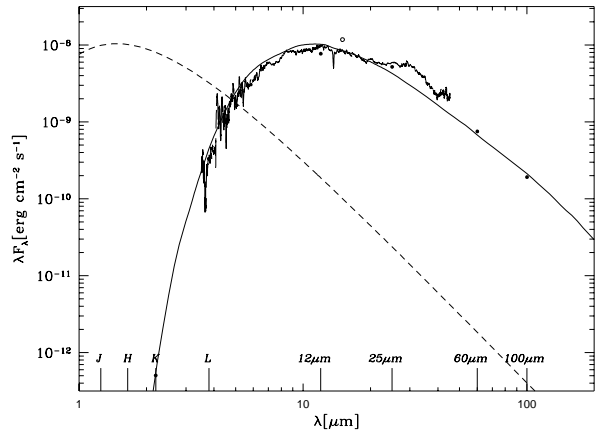}}
\caption{Observed and modelled spectral energy distribution.  The dot 
represents the near infrared and IRAS photometric results and the thin 
solid line is the ISO--SWS spectrum starting from band 1E. The flux at
15 micron measured from the ISO--CAM LW3 image is given by an open circle.
The thick line is calculated from the model with the values of the parameters 
listed in Table 1 and the long--dashed line represents the radiation 
spectrum from the central star which radiates as a blackbody of temperature 
2500\,K. }
\end{figure}

The model we used to fit the observational data is described in 
detail by Szczerba et al.  (1997).  In brief, the frequency--dependent 
radiative transfer equations are solved under the assumption of 
spherically--symmetric geometry simultaneously with the thermal balance 
equation for a dusty envelope.  The radiation of the central star is 
assumed to be a blackbody.  The mass loss rate is taken to be constant and 
the envelope is assumed to
expand at the velocity derived from the CO\,(2--1) line 
observation.  The dust opacity is represented by amorphous carbon grains of 
AC type (Bussoletti et al.  1987, Rouleau \& Martin 1991).

The appropriate values of important parameters for fitting the 
observational data are listed in Table 1.  The symbols have their
usualmeanings, but more details can be found in Szczerba et al.  (1997).  
Dynamical time t$_{dyn}$ means the time required for the matter to reach 
the outer radius of the envelope.

\begin{table}
\begin{flushleft}
\label{tablines}
\caption[]{Model parameters}
\begin{tabular}{ll} \hline
parameter & value \\ \hline

T$_{eff}$ & 2500\,K \\
L$_{star}$ & 8000\,L$_{\odot}$ \\
d & 4.25\,kpc \\

T$_{inner}$ & 760\,K \\
R$_{inner}$ & 5.8 $\times$ 10$^{14}$cm \\
R$_{out}$ & 7.0 $\times$ 10$^{17}$cm \\
\.{M} & 8.0 $\times$ 10$^{-5}$ M$_{\odot}$/yr \\
t$_{dyn}$ & 1.6 $\times$ 10$^{4}$ yr\\

Dust--to--gas ratio & 5.0 $\times$ 10$^{-3}$ \\
a$_{-}$ & 0.005\,$\mu$m \\
a$_{+}$ & 0.25$\,\mu$m \\
p (power--law index of size dist.) & 3.5\\ \hline
\end{tabular}

\end{flushleft}
\end{table}

The result from model calculation with the values listed in 
Table 1 is plotted in Fig.  1 by a heavy solid line, while the 
long--dashed line represents the radiation from the central star, 
which radiates
as a blackbody of effective temperature 2500\,K. The luminosity and 
distance of the star depend on each other, as well as the 
dust--to--gas mass ratio and mass loss rate assumed.  
We adopted the value of 8000 
L$_{\odot}$ for the luminosity.  The corresponding distance is 4.25 
kpc.  Since the outer radius R$_{out}$ is 7.0 $\times$ 10$^{17}$cm, 
the angular diameter of the circumstellar envelope is predicted to be
22 arcseconds at this 
distance.  This size is big enough to be resolved by the ISOCAM at the 
resolution mode of 1.5$''$.  However, the ISOCAM observations 
centered at 15\,$\mu$m are the most sensitive for the peak temperature 
of about 200\,K which, according to modeling results,
corresponds to radius of 4.1$\times$10$^{15}$\,cm 
and reflects a much smaller angular size of about 0.13 arcseconds.  
So the result of ISOCAM that no 
extension is seen may be attributed to a large distance of the object.  
The value of R$_{out}$ is determined from the model--fitting to the
observational results and its choice
affects mainly the flux intensity in the mid-- and far--infrared.
A reasoanable fit can be achieved with R$_{out}$ values ranging from 
5.0 $\times$ 10$^{17}$cm to 9.0 $\times$ 10$^{17}$cm assuming a constant mass
loss rate. For increasing mass loss rate (which means density 
distribution steeper than
r$^{-2}$), the outer radius should be larger to compensate for the smaller 
far infrared emission of the outer enevelope layers, 
but range of the allowed changes in
R$_{out}$ is quite similar. Note, however, such density distribution
cannot be much different from that corresponding to the constant mass loss 
rate due to strong constraints from the 60 and 100 $\mu$m flux densities
unless we assume
the dust optical properties at far infrared wavelengths have a 
less steep slope
than in the case of amorphous carbon used here.

The adopted dust--to--gas mass ratio of 5.0$\times$ 10$^{-3}$ corresponds to 
mass loss rate of 8.0 $\times$ 10$^{-5}$ M$_{\odot}$/yr. This mass 
loss rate is higher than 5.8 $\times$ 10$^{-6}$ M$_{\odot}$/yr 
deduced from CO\,(1--0) line (Loup et al. 1993) or 
1.4 $\times$ 10$^{-5}$ M$_{\odot}$/yr inferred from CO\,(2--1) line 
(Omont et al. 1993).  Note, in addition, that to get velocity of the shell 
around 14\,km/s dynamical considerations would suggest even smaller value of 
dust--to--gas mass ratio 
(see Steffen et al. 1998), and in consequence a larger
mass loss rate and a larger total mass of the envelope. However, 
the assumption of a
density distribution slightly more steep than r$^{-2}$ would cancel the 
increase in total mass of the envelope.
The calculation of the 
mass loss rate from the CO line suffers mainly from the uncertainty 
of the distance and mass fraction factor for CO molecules. For example,
the mass loss rate of W Hya derived from infrared water lines 
lies a factor about 30 above the estimates based on the CO line 
observation (Neufeld et al. 1996). The case of Y CVn is similar in that 
the mass 
loss rate derived from interpretation of far infrared ISOPHOT images 
is 2 orders of magnitude higher than that found 
from the CO line (Izumiura et al. 1996). Izumiura et al.(1996) explained 
the result for Y CVn by suggesting
that the far infrared and CO observations represent different epoches of mass 
loss. There may be another possibility that, the mass fraction of CO 
molecules is overestimated so that the mass loss rate is underestimated.  The 
mass loss rate of IRAS\,03313$+$6058 from modeling lies a little above that 
estimated from 
the flux at 60 micron, 4.8 $\times$ 10$^{-5}$ M$_{\odot}$/yr 
(Omont et al. 1993). Derivation of mass loss rate from the flux at IRAS 
bands depends on the distance and on the bolometric correction factor which
would induce some uncertainty.   
Though the mass loss rate derived from our modeling depends on the 
value of dust--to--gas ratio and density distribution, the result is 
relatively stable (probably better than a factor of two) because 
of the constraints required to fit the 
spectral energy distribution over the wide--range of wavelengths.   
The object then experiences 
quite strong wind, and has a circumstellar envelope perhaps as massive as 
1.282\,M$_{\odot}$.  By considering that the outer radius R$_{out}$ can 
vary in the range from 5.0 $\times$ 10$^{17}$ to 9.0 $\times$ 10$^{17}$cm,
the mass of the circumstellar envelope may be in the range of 0.92\,M$_{\odot}$
to 1.65\,M$_{\odot}$ under the assumed dust--to--gas ratio of 0.005. 
Since the mass of the cirsumstellar envelope appears to be 
above one solar mass,  
the star could very possibly be an intermediate--mass AGB star.

\subsection{The 30 micron emission}

As can be seen in Fig. 1, the object shows emission around 30 micron 
which is superimposed on the continuum radiation from the star and the 
circumstellar envelope.  The emission starts at 20 micron, peaks at 
about 30 micron and extends to longward of 40 micron.  Because of 
another unidentified emission around 43.6 micron (see Discussion), 
it is difficult to 
define the cut--off position of the 30 micron emission band from this 
spectrum, though the emission may extend to the long--wavelength limit 
of ISO--SWS. 

As described in the Introduction, MgS is regarded as a reasonable 
candidate to be the carrier of the 30 micron emission.  We tried to model 
this with the optical constants taken from the tables based on laboratory 
measurements of a MgS(90\%)--FeS(10\%)
mixture (Begemann et al.  1994). For our
computations,  we used two shapes for the grains, 
i.e.  the CDE (Continuous 
Distribution of Ellipsoids) and Mie theory (spherical 
grains).  Assumptions and method used 
 are described in detail by 
Szczerba et al.  (1997).

Because the temperature structure of the circumstellar envelope is 
determined from modeling of the spectral energy distribution, and because 
there is little difference in the dust temperature structure between the 
largest and smallest AC grains (a$_{-}$\,=\,0.005 and 
a$_{+}$\,=\,0.25\,$\mu$m) used in the calculations, 
the only free parameter after taking MgS 
into account is the number ratio between Sulfur atoms in MgS and 
total Hydrogen atoms  
n(S)/n(H).  Under the CDE approximation a value of 
3.0$\,\times$\,10$^{-6}$ for n(S)/n(H) gives a fit to 
the observed feature at the long--wavelength wing, though there is a 
little inadequacy in the short--wavelength wing of the band.  On the 
other hand, the Mie theory (spherical) grains can make up 
the emission at the short--wavelength wing.  This means that a 
combination of MgS grains shapes may account for the observed 
emission.  In Fig.  2, the observation and modeling of this band are 
shown.

\begin{figure}
\resizebox{\hsize}{!}{\includegraphics[angle=-90]{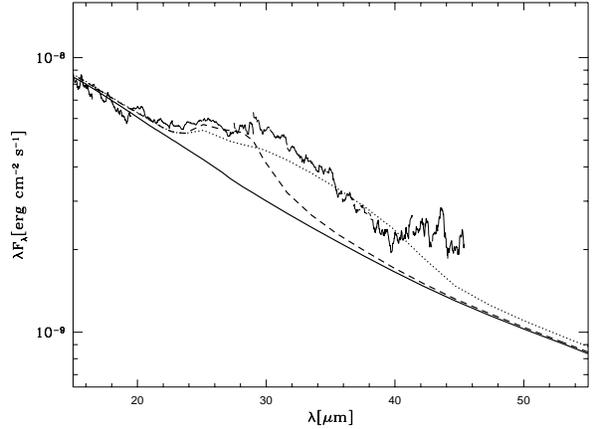}}
\caption{Fitting to the 30 micron emission.  The thin solid line is the 
ISO--SWS spectrum and the thick line stands for the model fitting 
result to spectral energy distribution, the same as in Fig.1.  The dot 
and dash lines represent the fitting by assuming the CDE and Mie 
approximation for the computation of optical properties of the MgS 
grains, respectively; n(S)/n(H) = 3.0\,$\times$\,10$^{-6}$. }
\end{figure}

\section{Discussion}
\indent

It is interesting to compare the 30 micron feature 
of this object with those of other AGB and post--AGB stars.  Fig.  3 exhibits 
the spectrum of
IRAS 03313$+$6058 together with the profiles of this feature for another 
AGB star, AFGL\,3068, and for the 
post--AGB object IRAS 22272$+$5435.  The data 
for AFGL\,3068 are taken from Yamamura et al. (1998) while the data for the 
IRAS\,22272$+$5435
are from Omont et al.  (1995).  Because these two objects are much 
brighter than IRAS\,03313$+$6058, their spectra are scaled downward to 
agree at about 20 $\mu$m with the spectrum  of IRAS\,03313$+$6058.  
Besides that IRAS\,22272$+$5435 has another feature at 21 micron, the most 
evident difference is that the emission of IRAS\,03313$+$6058 is much weaker 
than that of IRAS 22272$+$5435.  On the other hand,  AFGL\,3068 exhibits
a similar strength of the 30 $\mu$m band as does IRAS\,03313$+$6058 which
seems, however, to show more fine structures in this wavelength range. 

\begin{figure} 
\resizebox{\hsize}{!}{\includegraphics[angle=-90]{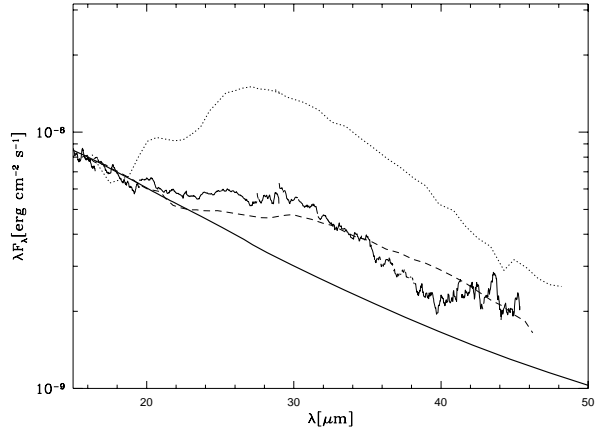}}
\caption[short title]{Comparison of the 30 micron band in 
IRAS\,03313$+$6058 (thin 
solid line) with the post--AGB object IRAS\,22272$+$5435 (dotted 
line) and  AGB star AFGL\,3068 (dashed line).  The modeling continuum 
spectrum at this band to IRAS 03313$+$6058 is represented by the thick 
solid line which lies below the emission features. }
\end{figure}

The 30 micron emission for IRAS\,22272$+$5435 was previously 
modeled  by wing of MgS grains  (Szczerba et al.  1997).  That fit 
resulted in an estimate of 
n(S)/n(H) = 4\,$\times$\,10$^{-6}$ for the highest dust 
temperature distribution and 1.6\,$\times$\,10$^{-5}$ for 
the lowest dust temperature distribution considered.  The value of  
3.0\,$\times$\,10$^{-6}$ for IRAS\,03313$+$6058 is very close to the 
case of highest dust temperature distribution of IRAS\,22272$+$5435.  
Then there is probably little difference in the sulfur abundance between these 
two objects and the strength of the 30 micron emission band may be 
influenced more strongly by other factors.  The temperature may be one of the 
important factors through the way that lower temperature favors the 
excitation of this band.  As the dust temperature of post--AGB envelopes is 
generally lower than that of AGB ones, this emission is 
stronger in post--AGB stars.  For example IRAS\,22272$+$5425 has colder 
dust than IRAS\,03313$+$6058 and much stronger emission at this band.  From 
laboratory experiment, MgS is one of the molecules which condenses at 
low temperature in the environment without oxides.  Up to now, 
this band emission is detected in the C--rich AGB stars only with 
cold and high optical depth dust shells. This 
may indicate the existence of a critical 
temperature to form the 30 micron emission band carrier, and that such low 
temperatures together with approporiate chemical conditions are only 
possible in the 
extreme carbon stars. On the other hand, it is still not clear if carbon stars
with smaller mass loss rates could form a carrier of this band. The existence 
of the 30 $\mu$m emission in AFGL\,2155 and IRC$+$40540 (Yamamura et al. 1998),
neither of them was classified as extreme C--stars by Volk et al. (1992), 
suggests that formation of the approporiate chemical material is much more
common and that higher envelope temperature in other carbon stars (not 
extreme ones)
does not allow us 
to detect this emission feature. But it could be as well that some special
chemical reactions responsible for the carrier formation of the 30 $\mu$m 
band are efficient only when temperature is enough low and/or the chemical 
composition is approporiate. Unfortunately, since without optical spectra 
the determination of the chemical abbundances 
is rather impossible other methods of
investigation should be elaborated, and especially better statistics created
by the ISO data could help solve the problem of the 30 $\mu$m carrier 
formation. 

Besides this feature around 30 micron, some other features, e.g. 
absorptions around 7.5 and  14 micron, emissions around 41 and 43.5 $\mu$m
in the SWS spectrum of this object are not discussed here; they are currently
under investigation. We note only that the absorptions are probably related
to the C$_2$H$_2$ and/or HCN molecular bands, while emissions could be related
to the crystaline silicates (especially as the 
enstatite mass absorption coefficient
matches these emissions well - see J{\"a}ger et al. 1998). If crystalline
silicate emissions are confirmed then it will allow to deduce an 
evolutionary status of IRAS\,03313$+$6058 which could bring some more 
information on the exciting transition phase between oxygen and carbon rich
parts of the AGB evolution.

\acknowledgements{B.W.J. thanks the people in N. Copernicus 
Astronomical Center, Torun for their help and support.  We express also our
gratitudes to Dr. Kevin Volk for his careful reading of the 
manuscript and useful suggestions. This work has been 
partly supported by grant 2.P03D.002.13 of the Polish State Committee 
for Scientific Research.}


\begin{thebibliography}{}

\bibitem{} Begemann B., Dorschner J., Henning T., Mutschke H., Thamm E., 
1994, ApJ 423, L71 

\bibitem{}Bussoletti E., Colangeli L., Borghesi A., Orofino V., 
1987, A\&AS 70, 257 

\bibitem{}Cesarsky C.J., Abergel A., Agnese P., Altieri B., Augueres J.L., et 
al., 1996, A\&A 315, L32 

\bibitem{} Chan K.W., Moseley S., Casey S., et al., 1997, ApJ 483, 798 

\bibitem{} de Graauw Th., Haser L., Beintema D., et al.,  1996, A\&A 315, L345 

\bibitem{} Forrest W.J., Houck J.R., McCarthy J.F., 1981, ApJ 248, 195 

\bibitem{} Galt J., Sun K., Frankow J., 1989, AJ 98, 2182 

\bibitem{} Goebel J., Moseley S., 1985, ApJ 290, L35 

\bibitem{} Izumiura H., Hashimoto O., Kawara K., Yamamura I., 
Waters L.B.F.M., 1996, A\&A 315, L221

\bibitem{} J{\"a}ger C., Molster F.J., Dorschner J., et al., 
1998, A\&A 339, 904

\bibitem{} Jiang B.W., Hu J.Y., 1992, ChA\&A 16, 416 

 \bibitem{} Jiang B.W., Deguchi S., et al., 1996, ApJS 106, 463 

\bibitem{} Jiang B.W., Deguchi S., Hu J.Y., et al., 1997, AJ 113,  1315

\bibitem{} Kwok S., Volk K., Hrivnak B.J., 1989, ApJ 345, L51 

\bibitem{} Kwok S., Volk K., Bidelman W.P. 1997, ApJS 112, 557 

\bibitem{} Le Squeren A.M., Sivagnanam P., Dennefeld M., David P., 
1992, A\&A 254, 133 

\bibitem{} Loup C., Forveille T., Omont A., Paul J.F., 1993, A\&AS 99, 291 

\bibitem{} Low F., Rieke G., Armstrong K., 1973, ApJ 183, L105 

\bibitem{} Neufeld D., Chen W., Melnick G., et al., 1996, A\&A 315, L237

\bibitem{} Omont A., Loup C., Forveille T., te Lintel Hekkert P., 
Habing H. \& Sivagnanam P. 1993, A\&A 267, 515 

\bibitem{} Omont A., Moseley S., Cox P., et al., 1995, ApJ 454, 819
 
\bibitem{} Rouleau F., Martin P.G. 1991, ApJ 377, 526 

\bibitem{} Steffen M, Szczerba R., Sch{\"o}nberner D., 1998, A\&A 337, 149
 
\bibitem{} Szczerba R., Omont A., Volk K., Cox P., Kwok S., 
1997, A\&A 317, 859 

\bibitem{} Volk K., Kwok S., Langill P., 1992, ApJ 391, 285 

\bibitem{} Volk K., Kwok S., Woodsworth A., 1993, ApJ 402, 292 

\bibitem{} Wouterloot J., Brand J., Fiegle K., 1993, A\&AS 98, 589

\bibitem{} Yamamura I., de Jong T., Justtanont K., Cami J., 
Waters L.B.F.M., 1998, Ap\&SS 255, 351

\end{thebibliography}
\end{document}